\documentclass[referee]{raa}            

\usepackage{graphicx,times}             
\usepackage{natbib}
\usepackage{amssymb,amsmath}
\usepackage{upgreek}
\bibpunct{(}{)}{;}{a}{}{,}

\begin{document}

   \title{A new post-hoc flat field measurement method for the Solar X-ray and Extreme Ultraviolet Imager onboard the Fengyun-3E satellite
}

   \volnopage{Vol.0 (20xx) No.0, 000--000}      
   \setcounter{page}{1}          

   \author{Qiao Song\inst{1,2,7}, Xianyong  Bai \inst{3,*}, Bo Chen \inst{4,*}, Xiuqing Hu \inst{5,7,*}, Yajie Chen \inst{6}, Zhenyong Hou \inst{6}, Xiaofan Zhang \inst{3},  Lingping He \inst{4}, Kefei Song \inst{4}, Peng Zhang \inst{5,7}, Jing-Song Wang \inst{1,7}, Xiaoxin Zhang \inst{1,7}, Weiguo Zong\inst{1,7}, Jinping Dun \inst{1,7}, Hui Tian \inst{6}, Yuanyong Deng \inst{3}
   }

   \institute{Key Laboratory of Space Weather, National Satellite Meteorological Center (National Center for Space Weather), China Meteorological Administration, Beijing, 100081, People's Republic of China\\
   \and
   State Key Laboratory of Space Weather, Chinese Academy of Sciences, Beijing, 100190, People's Republic of China \\
   \and
   National Astronomical Observatories, Chinese Academy of Sciences, Beijing, 100012, People's Republic of China; {\it xybai@bao.ac.cn}\\
    \and
   Changchun Institute of Optics, Fine Mechanics and Physics, Chinese Academy of Sciences, Changchun, 130033, People's Republic of China; {\it chenb@ciomp.ac.cn} \\
   \and   
   Key Laboratory of Radiometric Calibration and Validation for Environmental Satellites, National Satellite Meteorological Center (National Center for Space Weather), China Meteorological Administration, Beijing, 100081, People's Republic of China; {\it huxq@cma.gov.cn} \\
   \and
   School of Earth and Space Sciences, Peking University, Beijing, 100871, People's Republic of China\\
    \and 
   Innovation Center for FengYun Meteorological Satellite, China Meteorological Administration, Beijing, 100081, People's Republic of China}

\vs\no
   {\small Received~~20xx month day; accepted~~20xx~~month day}

\abstract{The extreme ultraviolet (EUV) observations are widely used in solar activity research and space weather forecasting since they can observe both the solar eruptions and the source regions of the solar wind. Flat field processing is indispensable to remove the instrumental non-uniformity of a solar EUV imager in producing high-quality scientific data from original observed data. Fengyun-3E (FY-3E) is a meteorological satellite operated in Sun-synchronous orbit, and the routine EUV imaging data from the Solar X-ray and Extreme Ultraviolet Imager (X-EUVI) onboard FY-3E has the characteristics of concentric rotation. Taking advantage of the concentric rotation, we propose a post-hoc flat field measurement method for its EUV 195 {\AA} channel in this paper. This method removes small-scale and time-varying component of the coronal activities by taking the median value for each pixel along the time axis of a concentric rotation data cube, and then derives large-scale and invariable component of the quiet coronal radiation, and finally generates a flat field image. The flat field can be generated with cadences from hundreds of minutes (one orbit) to several days. Higher flat field accuracy can be achieved by employing more data. Further analysis shows that our method is able to measure the instrumental spot-like non-uniformity possibly caused by contamination on the detector, which mostly disappears after the in-orbit self-cleaning process. It can also measure the quasi-periodic grid-like non-uniformity, possibly from the obscuration of the support mesh on the rear filter. After flat field correction, these instrumental non-uniformities from the original data are effectively removed. Moreover, the X-EUVI 195 {\AA} data after dark and flat field corrections are consistent with the 193 {\AA} imaging data from the Atmospheric Imaging Assembly onboard the Solar Dynamics Observatory, verifying the suitability of the method. The post-hoc method does not occupy observation time, which is valuable for space weather operations. Our method is not only suitable for FY-3E/X-EUVI but also a candidate method for the flat field measurement of future solar EUV telescopes.
}

   \authorrunning{Q. Song et. al.}            
   \titlerunning{New flat field method for FY-3E solar EUV images}  

   \maketitle

%
%
\section{Introduction}
Space weather is important to satellites, spacecraft and astronauts as human exploration expands from our home planet Earth to the Moon, Mars, and beyond. The expansion of technological infrastructure on Earth also highlights the impact of space weather. The primary driver of space weather is the Sun \citep{2006LRSP....3....2S}. Its highly dynamic atmosphere produces various activities such as solar flares and coronal mass ejections. In the extreme ultraviolet (EUV) wavelength, almost all of solar activities in the outer layers of the solar atmosphere (i.e., the chromosphere, the transition region, and the corona) can be observed. Solar active regions usually show bright loop systems in EUV images, and with the evolution of the magnetic field in active regions, the loop systems of the active regions also change dynamically. When a flare occurs, the source region can be identified from EUV images, and then the flare process can be tracked and studied \citep{Song2022}. In EUV images, coronal holes are also observed, which are cooler, less dense regions in solar atmosphere. Coronal holes are generally believed to be the source of the solar wind that can cause geomagnetic storms and other space weather disturbances \citep{2007LRSP....4....1P}. Other solar activities, such as EUV waves \citep{2022ApJ...928...98H} and filament eruptions can also be well represented on EUV images. 
Moreover, the variability of solar EUV radiation influences Earth's atmosphere and further climate on different time-scales. On the short time-scale, major solar flares can affect Earth's ionosphere, causing disturbances such as blackout of high-frequency radio communication and degradation of low-frequency navigation signals. Solar influence on Earth's climate is often reflected in the radiation changes of EUV and other wavelengths on the 11-year period of solar cycles \citep{Lockwood2012}.

Therefore, the EUV wavelength is widely used in solar activity research and space weather forecasting. Due to the absorption of Earth's atmosphere, solar EUV telescopes need to be onboard a satellite or spacecraft. For example, the Transition Region and Coronal Explorer (TRACE, \citealt{Handy1999}), Extreme ultraviolet Imaging Telescope (EIT, \citealt{1995SoPh..162..291D}) for the SOHO Mission, the Sun Earth Connection Coronal and Heliospheric Investigation (SECCHI, \citealt{2008SSRv..136...67H}) on the twin STEREO spacecraft, the Sun Watcher using Active Pixel System detector and Image Processing (SWAP, \citealt{2013SoPh..286...67H}) telescope on the PROBA2 technological mission, the Atmospheric Imaging Assembly (AIA, \citealt{2012SoPh..275...17L, Boerner2012}) of the Solar Dynamics Observatory (SDO, \citealt{2012SoPh..275....3P}), and the Solar X-ray and Extreme Ultraviolet Imager (X-EUVI, \citealt{chen2022, chen2022b}) onboard the Fengyun-3E (FY-3E, \citealt{zhang2022}) satellite. FY-3E is the latest one of China's second generation polar-orbiting meteorological satellites. Launched on July 5, 2021, FY-3E is now operated in a Sun-synchronous orbit (i.e., dawn/dusk orbit) at $\sim$836 km altitude with an inclination of 98.75{\degr}. The X-EUVI has multiple X-ray and EUV channels that can be used for space weather forecasting and scientific research. In the EUV wavelength, it provides full-disk 195 {\AA} images of the Sun with a moderate spatial resolution ($\sim$2.{\arcsec}5 per pixel) and a high time resolution up to a few seconds.
In addition, the solar images of X-EUVI can be combined with the images of the Advanced Space-based Solar Observatory (ASO-S, \citealt{2019RAA....19..156G}) and the Chinese H${\alpha}$ Solar Explorer mission (CHASE, \citealt{Li2022}) for multi-band comprehensive observations to study solar activities.

Due to the large-scale vignetting in the optical system, the small-scale pixel-to-pixel variation of the charge-coupled device (CCD) detector and the unavoidable intermediate-scale contamination or dust from the optical elements and the CCD detector, the recorded data from all of the EUV solar telescopes is non-uniform even if the target is a light source with uniform light emitting area. In addition, the nickel mesh supporting the focal-plane filter can also create grid-like features on the observed image, resulting in increased non-uniformity \citep{Boerner2012, swap2013}. Therefore, flat field processing is generally employed to remove the non-uniformity of the whole system in the production of high-quality scientific data from original observed data. However, up to now, how to acquire precise flat field image is still one of the most challenging problems for full-disk solar observations. The best and easy way to measure the instrumental flat field is to find a highly uniform light source \citep{Howard2008,Kentischer2008,bai2017,Wang2019}, which is extremely difficult at the EUV wavelength. An alternative way is to adopt a non-uniform light source such as the Sun itself and the flat field is generated by taking several offset solar images with slightly shifted locations on the detector. The offset can be realized by changing the tip-tilt angles of the second mirror. It can also be taken by changing the pointing of the spacecraft. Both of the methods are used for flat field measurement of the EUV imagers of the Sun, e.g., TRACE and SDO/AIA. Once series of offset images are taken, the shifts in the x and y directions relative to the reference image are determined and then the flat field is generated by the Kuhn-Lin algorithm \citep{Kuhn1991}. Other algorithms were also proposed to derive flat field \citep{Chae2004,Xu2016,Gao2020,Li2021}.

If the offset solar images have acentric rotation on the detector due to the attitude control of a spacecraft, there are both rotation and shift during the image registration process, making it difficult to employ the above-mentioned methods to generate the flat field. In this paper, we propose a new post-hoc method to create flat field from the routine concentric rotated coronal images taken by FY-3E/X-EUVI. Similar method has been used for the generation of flat field and polarized offset from regular observed full-disk photospheric filtergrams and Stokes $\frac{Q}{I},\ \frac{U}{I},\  \frac{V}{I} $ maps, respectively \citep{Potts2008,Wachter2009,bai2018}. Here we extend it to regular observational data of the corona. The principle of our method is presented in Section 2. Section 3 shows the results of the method application in observational images, following with a conclusion and discussion section (Section 4).

\section{Principle of the post-hoc flat field measurement method}

The recorded image I(x,y) taken by a solar EUV telescope can be expressed as:

\begin{equation}\label{eq1}
  I(x,y)=I_{s}(x,y)\times f(x,y)+I_{d},
\end{equation}

where $I_{s}(x,y), \ f(x,y)$ and $I_{d}$ represent the solar-disk coronal radiation maps, flat field and dark field of the whole system, respectively. 
From Eq. \ref{eq1}, the solar-disk radiation $I_{s}(x,y)$ can be recovered from the original recorded data $I(x,y)$ once the flat field and dark field are obtained. The key step is to measure the flat field and dark field. If we try to divide the solar-disk coronal radiation into two parts, $I_{s}(x,y)$ becomes:

\begin{equation}\label{eq3}
I_{s}(x,y)=I^{l}_{s}(x,y)+I^{h}_{s}(x,y).
\end{equation}

Here $I^{l}_{s}(x,y)$ represents the large-scale component that is to say the quiet coronal radiation with limb brightening pattern caused by the optical thin effect of EUV lines \citep{1970SoPh...11...42W}, which rarely changes with time (i.e., with low frequency in time variation). $I^{h}_{s}(x,y)$ represents the time-varying medium and small-scale components, such as the active region loops, the coronal holes, and the coronal bright points (i.e., with high frequency in time variation).

If a series of EUV coronal radiation maps $I_{s}(x,y,t)$ having concentric rotation within several days are used to create a data cube and the median value is taken for each spatial pixel, the time-varying component $[I^{h}_{s}(x,y,t)]_{median} \approx 0$ while the large-scale component remains almost the same. Then Eq. \ref{eq3} can be written as:

\begin{equation}\label{eq4}
[I_{s}(x,y,t)]_{median} \approx [I^{l}_{s}(x,y,t)]_{median}+[I^{h}_{s}(x,y,t)]_{median}=I^{l}_{s}(x,y).
\end{equation}

Once the non-uniformity of the $I^{l}_{s}(x,y)$ is known, we can divided it from the $[I_{s}(x,y,t)]_{median}$ image and the final image can be treated as a uniformed light source used for flat field measurement. According to Eq. \ref{eq1}, the flat field is calculated from the median of dark-corrected routine EUV coronal images with :

\begin{equation}\label{eq5}
f(x,y)=\frac{[I(x,y,t)-I_{d}]_{median}}{I^{l}_{s}(x,y)}.
\end{equation}

Now the main task is to calculate the the large-scale components $I^{l}_{s}(x,y)$, which has circular symmetry in general since what we observed is a projection of the spherical coronal radiation. It is more convenient to make a conversion of coordinates from Cartesian lattice to polar coordinates to deal with the circular symmetry.

\begin{equation}\label{eq6}
I^{'}(r,\theta)=I^{'}(x,y)=[I(x,y,t)-I_{d}]_{median}.
\end{equation}

The r and $\theta$ in Eq. \ref{eq6} are the radius and azimuth angles, respectively. And $I^{'}(r,\theta)$ is the median of series of dark-corrected routine coronal images in polar coordinates. If we take the median value for the intensities along all of the azimuth angles with the same radius, the quiet coronal radiation and limb brightening curve can be derived, which is described in the below equation:

\begin{equation}\label{eq7}
I^{'}(r)=I^{'}(r,\theta)|_{median\ along\  \theta \ direction}.
\end{equation}

We are able to further obtain the large-scale coronal component $I^{l}_{s}(x,y)$ by filling in the values in the azimuth direction for each radius with the values in Eq. \ref{eq7} and transforming it back to Cartesian coordinates. Finally, the flat field is generated from Eq. \ref{eq5}, and the flat field corrected corona radiation is derived with Eq. \ref{eq1}.

\section{Results}

We use the EUV coronal images from FY-3E/X-EUVI 195 {\AA} channel to test the performance of the method. Since FY-3E is a polar-orbiting meteorological satellite, its main pointing is Earth and its observational images of the Sun have significant rotation. The X-EUVI has a two-dimensional turntable to point and track the Sun. It also equipped with an imaging stabilization system to rapidly compensate the jitters from the satellite and the turntable to improve the spatial resolution. The routine coronal images rotate with an orbit period of $\sim$101.5 minutes, which is concentric with the help of the imaging stabilization system. If we create a data cube from the staked concentric rotated images taken within one orbit and take the median values for each spatial pixel, the assumption that the $I^{h}_{s}(x,y) \approx 0$ still works. Then we can employ the method in section 2 to derive the flat field.

\subsection{Flat field generated from X-EUVI's data}

\begin{figure}
\includegraphics[bb=0 190 592 592,width=1\textwidth,clip]{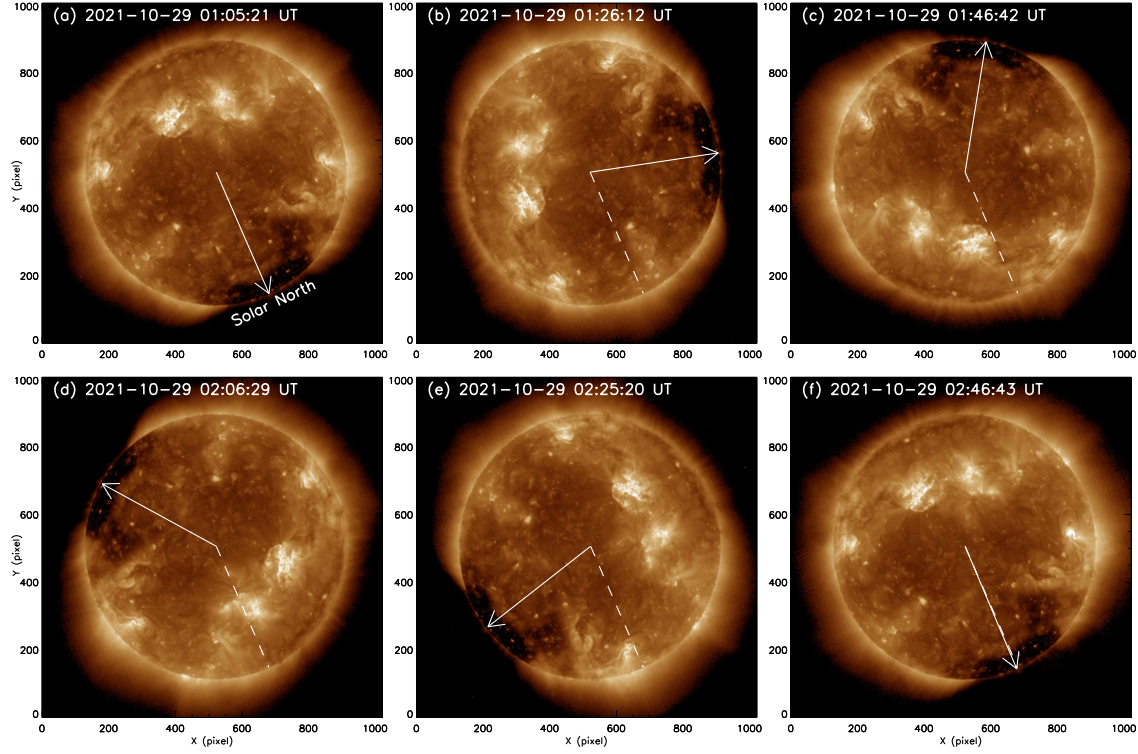}
\caption{Time-series of the X-EUVI 195 {\AA} images within one orbit showing counterclockwise rotation. The arrows in the figure indicate the north pole direction of the Sun, and the dashed lines indicate the initial position.}
\label{fig:fig1}
\end{figure}

Figure \ref{fig:fig1} shows the time-series of the EUV 195 {\AA} images observed in the first orbit on October 29, 2021, with an interval of about 20 minutes between each panel of the figure. The exposure time of each image is 0.8 second and the dark field has been corrected. The dark field of X-EUVI is taken in the calibration mode with same time as normal exposure, but with the shutter closed, and the telescope is pointed in the same way as the normal observation, so as to ensure that the dark field and solar observation images have the same temperature conditions on the telescope and the CCD detector. As can be seen from the figure, there is a polar coronal hole near the north pole of the Sun. The arrows in the figure indicate the changing north pole direction, and the dashed lines indicate the initial position of concentric image rotation. The horizontal and vertical coordinates of the image are pixels. As the satellite orbits the Earth, the solar image rotates counterclockwise at an angular velocity of about 3.5{\degr} per minute. At 01:05 UT, the coronal hole is located at the lower right of the image (Figure \ref{fig:fig1}(a)), and it turns counterclockwise to the upper position of the image (Figure \ref{fig:fig1}(c)) at 01:46 UT. Finally, after the image is rotated nearly 360{\degr}, the coronal hole switches back to the lower right of the image again at 02:46 UT (Figure \ref{fig:fig1}(f)).

\begin{figure}
\includegraphics[bb=32 30 592 592,width=1\textwidth,clip]{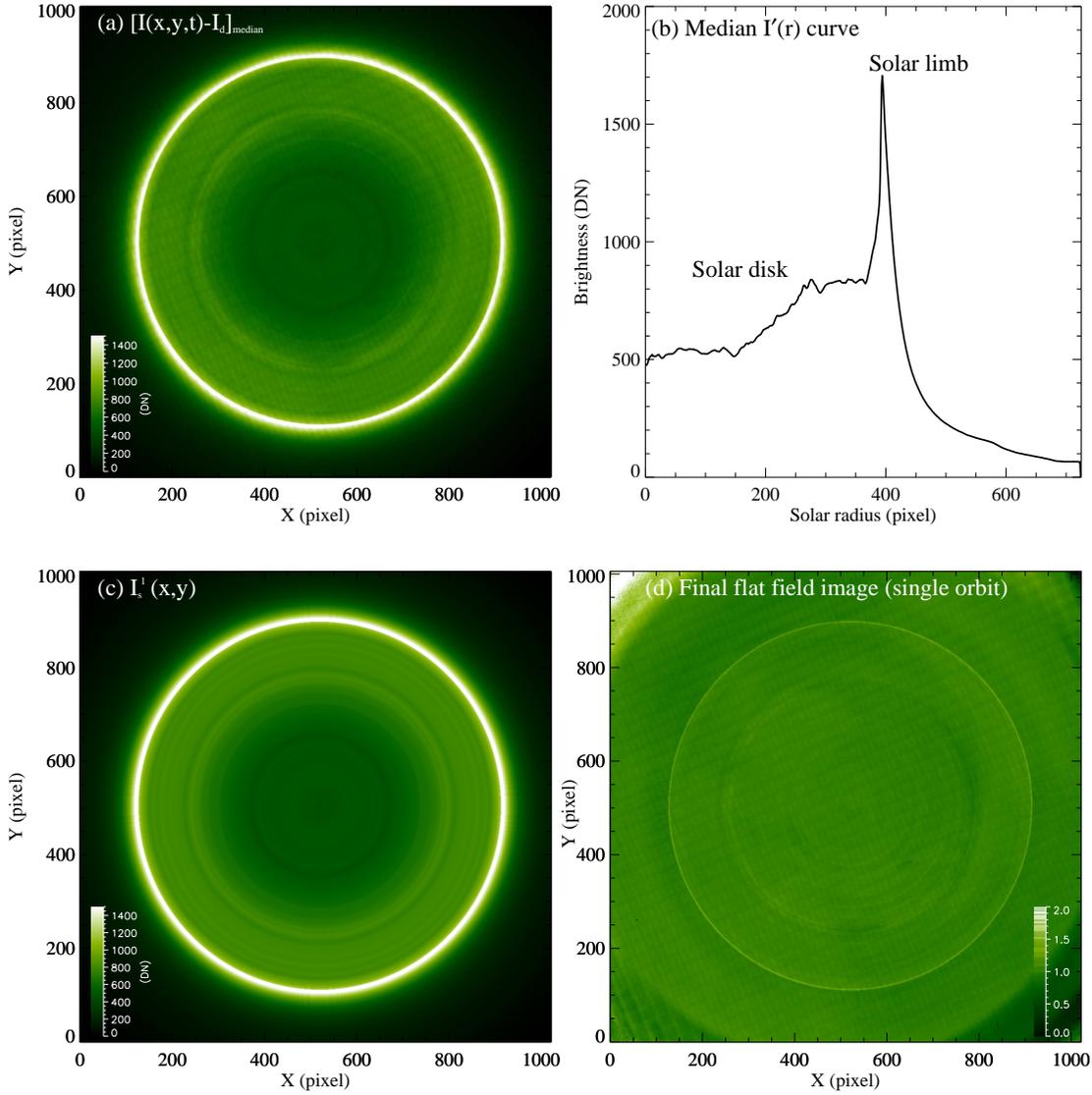}
\caption{Calculation and generation process of the flat field image. (a) is the dark corrected median data $[I(x,y,t)-I_{d}]_{median}$ from time series images of the first orbit on October 29, 2021, and (b) shows the median $I^{'}(r)$ curve representing the quiet coronal radiation derived from panel (a). (c) displays the quiet coronal radiation $I^{l}_{s}(x,y)$ image derived from panel (b), and (d) is final flat field image of the time series data in Cartesian coordinates.}
\label{fig:fig2}
\end{figure}

With the first orbit data on October 29, 2021 shown in Figure \ref{fig:fig1}, we derived the flat field using the method described in Section 2 and the result is presented in Figure \ref{fig:fig2}. Figure \ref{fig:fig2}(a) is the $[I(x,y,t)-I_{d}]_{median}$ image in the Cartesian coordinate system obtained by the median value of each pixel from a time series of 417 frames. After the processing, the small-scale components of the coronal radiation $I^{h}_{s}(x,y)$ is almost eliminated. From the image, we can see the residual large-scale coronal component $I^{l}_{s}(x,y)$, i.e., the quiet radiation with circular symmetry and the limb brightening pattern. We can also see the instrumental non-uniformity, i.e., the spot-like features and quasi-periodic grid-like features. Figure \ref{fig:fig2}(b) is the median $I^{'}(r)$ curve calculated from Eq. \ref{eq7} after transforming Figure \ref{fig:fig2}(a) to polar coordinates with Eq. \ref{eq6}. Since solar active regions and other active features have been smoothed out, we can see the quiet coronal radiation from solar disk and the bright solar limb. Figure \ref{fig:fig2}(c) displays the quiet coronal radiation $I^{l}_{s}(x,y)$ image derived from Figure \ref{fig:fig2}(b), and it can be seen that only bright limb and concentric circle features remain on this image. Figure \ref{fig:fig2}(d) is the final flat field image obtained according to Eq. \ref{eq5}. At this time, there are mostly instrumental non-uniformity left, such as the grid-like features and some inhomogeneous intermediate-scale features.

\begin{figure}
\includegraphics[bb=48 30 592 592,width=1\textwidth,clip]{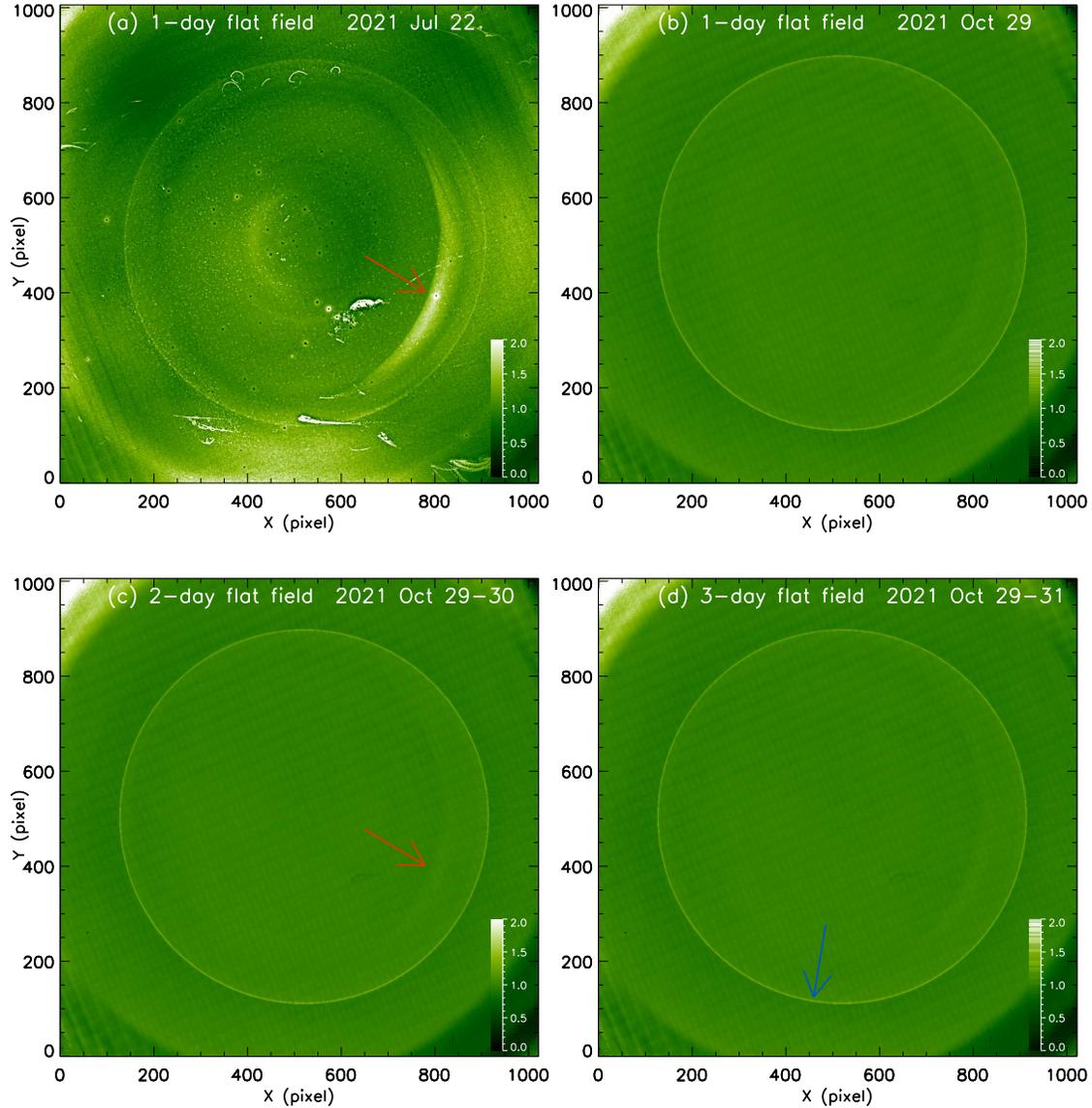}
\caption{Flat field images of one-day, two-day and three-day data, respectively. The red arrows in panels (a) and (c) indicate the arc-shaped pattern that possibly caused by the vignetting effect from the optical system. The blue arrow in panel (d) shows the narrow ring at solar limb which possibly caused by limited instrument pointing accuracy and stability.}
\label{fig:fig3}
\end{figure}

In Figure 2, the flat field is generated with a data cube in one orbital period. The accuracy of the flat field depends on the elimination of small-scale components of the corona $I^{h}_{s}(x,y)$. The discussion from \citet{Potts2008} and \citet{bai2018} indicate that the accuracy of the generated flat field is directly proportional to $\frac{\delta I}{\sqrt{n}}$, where $n$ is the number of the employed images and $\delta I$ is the fluctuation of the radiation among various coronal features. Small $\delta I$ is satisfied by choosing the observed EUV images with less coronal hole or solar active regions. Moreover, more EUV images with larger $n$ helps to improve the accuracy, which is confirmed in Figure \ref{fig:fig3}. Figure \ref{fig:fig3}(a) and \ref{fig:fig3}(b) are flat field images produced using one-day's data of July 22 and October 29, 2021, respectively.
These two sets of data represent two typical instrumental non-uniformity of the X-EUVI 195 {\AA} channel. Obscuration of the Earth and the satellite itself, as well as multi-channel observation modes, can reduce the number of available EUV images in one orbit for the flat field processing, and then affect the accuracy of the flat field image.
Therefore, the available EUV images in October 29, 2021 are more than twice that of the data in July 22, 2021, which results in a better flat field accuracy in October 2021. Figures \ref{fig:fig3}(b) and \ref{fig:fig3}(c) are the flat field images generated from the data of one day on October 29 (5894 frames) and two days of October 29-30 (11928 frames), respectively. Compared Figure \ref{fig:fig2}(d) with Figures \ref{fig:fig3}(b) and \ref{fig:fig3}(c), the residual inhomogeneity of the time-varying component $I^{h}_{s}(x,y)$ is found in the flat field using single-orbit data in Figure \ref{fig:fig2}(d). With the increasing frames $n$, the residual solar-disk features on the flat field images becomes less and less. From the flat field using three-day's routine EUV images (October 29-31, with 17948 frames) shown in Figure \ref{fig:fig3}(d), the residual solar features basically disappear, and the grids and spot-like features are more prominent. The arc-shaped pattern (see the red arrows) in the right part of Figures \ref{fig:fig3}(a) and \ref{fig:fig3}(c) possibly caused by the vignetting effect from the optical system of the instrument. It should be pointed out that the errors at the edge of the flat field image, especially at the four corners, may be large, due to the small number of data pixels there.

\subsection{Evaluation of the flat field processing}

\begin{figure}
\includegraphics[bb=32 30 592 592,width=1\textwidth,clip]{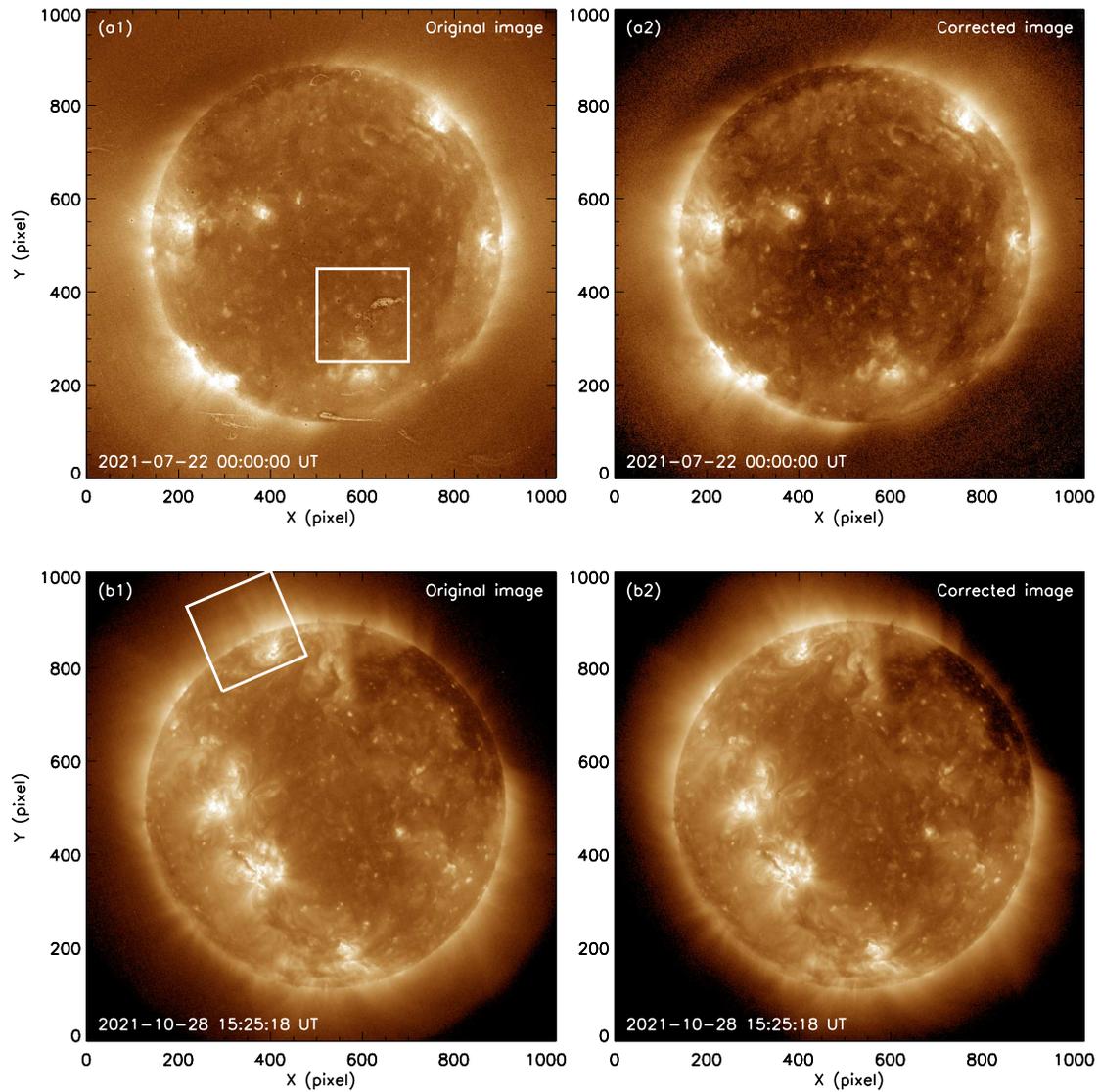}
\caption{The solar full-disk images before (left column) and after (right column) the flat field processing for July 22 (upper row) and October 28 (bottom row), 2021, respectively. The boxes in (a1) and (b1) represent the field of view of Figure \ref{fig:fig5}(a1) and Figure \ref{fig:fig5}(b1), respectively.}
\label{fig:fig4}
\end{figure}

\begin{figure}
\includegraphics[bb=0 190 592 592,width=1\textwidth,clip]{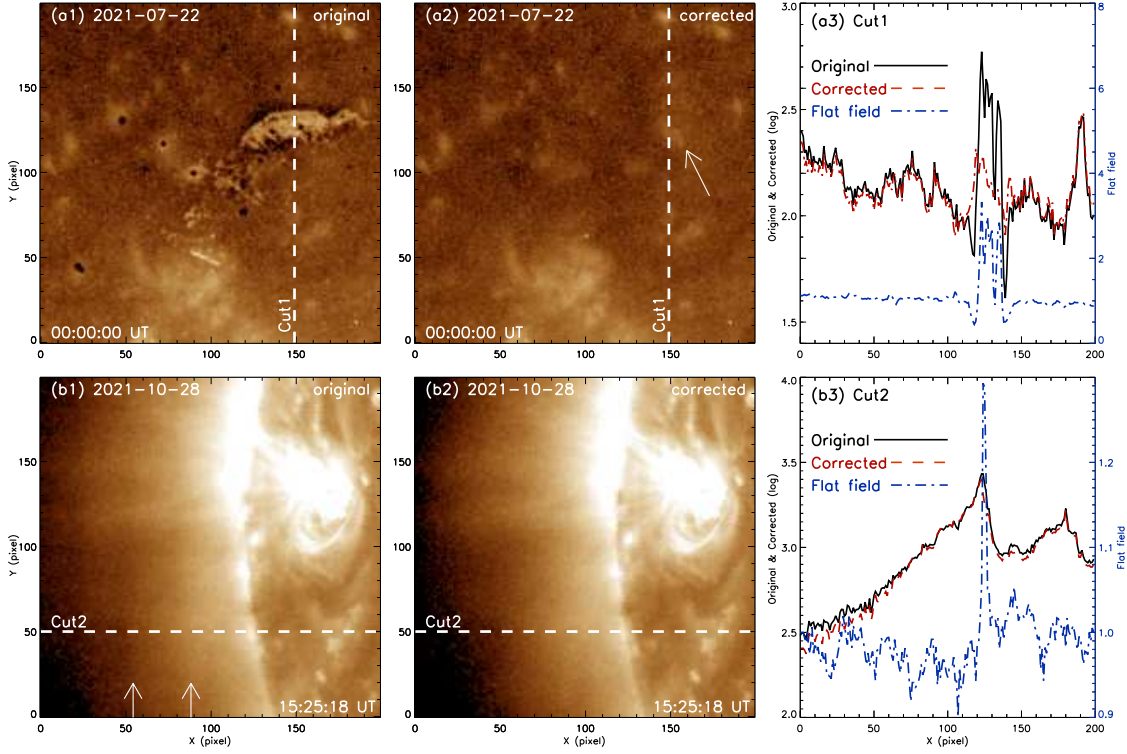}
\caption{The local region from Figure \ref{fig:fig4} before (left column) and after (middle column) the flat field processing. The solid, red dashed, blue dot-dashed lines in the right column correspond to the original intensity, the intensity after flat field processing, and the values used in the flat field extracted from the dashed lines (i.e., Cut1 and Cut2) in the left and middle columns. The arrow in panel (a2) indicates a coronal bright point that revealed after the flat field processing. The two arrows in panel (b1) mark the positions of two quasi-periodic grid-like features.}
\label{fig:fig5}
\end{figure}

With the flat field calculated from one day's data, we carried out the flat field processing of the original EUV 195 {\AA} images according to Eq. \ref{eq1} for two days in July 2021 and October 2021, respectively. The results are arranged in Figure \ref{fig:fig4} and Figure \ref{fig:fig5}. Figures \ref{fig:fig4}(a1) and \ref{fig:fig4}(a2) are the original full-disk data and the calibrated image after dark and flat field corrections, respectively. Figures \ref{fig:fig5}(a1) and \ref{fig:fig5}(a2) are for a local region in order to better show the improvement after the correction of instrumental non-uniformity. On July 22, X-EUVI's self-cleaning mode is about to start, and there are multiple spot-like features on the image (see Figure \ref{fig:fig4}(a1)) possibly caused by the adsorbed contaminant on the CCD detector. It can be seen that these spot-like features basically disappear after the flat field correction (Figure \ref{fig:fig4}(a2)). In Figure \ref{fig:fig5}(a1), we extracted one line (Cut1) at x=150 with a spot-like feature from y=115 to y=145. From the flat field at the positions of Cut 1 marked by the blue dot-dashed line in Figure \ref{fig:fig5}(a3), the normalized values in the spot-like features are in the range of 0.5 to 3. The coronal bright point (see the arrow in Figure \ref{fig:fig5}(a2) and the red dashed line in Figure \ref{fig:fig5}(a3)), which is hardly found at the position of the spot-like feature, is clearly seen after the flat field correction. According to the flat field data, the contamination on the detector on July 22, 2021 is the most serious so far, and the successful correction of heavily contaminated images demonstrates the effectiveness of our method.

Figure \ref{fig:fig4}(b1) and Figure \ref{fig:fig4}(b2)  are the original and corrected images on October 28, 2021, respectively. A self-cleaning of the instrument has just been completed on this day, so there are no obvious spot-like features from the deposited contaminant. From the flat field derived from one day's concentric rotation data in Figure \ref{fig:fig3}(b) on October 29, we can also confirm that the instrumental non-uniformity from the contaminant is non-significant. It indicates that the self-cleaning mode of X-EUVI is quite efficient to reduce the non-uniformity caused by the contaminant. After self-cleaning, the most obvious instrumental non-uniformity is the small grid-like features caused by the differential obscuration from the support mesh of the X-EUVI's focal-plane filter (i.e., the rear filter) seen in Figure \ref{fig:fig3}(b). 
The support mesh is installed at a distance of 16.7 mm from the CCD detector, with a grid period of 363 $\upmu$m and a grid width of 30 $\upmu$m. In Figure \ref{fig:fig5}(b1), we mark the positions of two grid-like features with arrows. The grids become vertical after we rotate the images in Figure \ref{fig:fig4}(b1) anti-clockwise with an angle of 67{\degr}. A local region near the solar limb (see the box in Figure \ref{fig:fig4}(b1)) is selected and shown in Figure \ref{fig:fig5}(b1), and Figure \ref{fig:fig5}(b2) is the same region after the flat field correction. Compare the two images before and after flat field correction, the grid-like features almost disappeared (Figure \ref{fig:fig4}(b2) and Figure \ref{fig:fig5}(b2)). A horizontal line in Figure \ref{fig:fig5}(b1) is selected for a slice of Cut2. The original, flat field and flat field corrected values in the line are presented in Figure \ref{fig:fig5}(b3). From the blue dot-dashed flat field values, the grid-like features have fluctuation with an amplitude of 14\% (from 0.91 to 1.05) and a period of about 30 pixels. The pixel size of the CCD detector of X-EUVI is 13 $\upmu$m, so the period of the grid-like features is about 390 $\upmu$m, which does not exactly equal to the period of the support mesh. However, considering the mesh locates at a distance of 16.7mm from the CCD detector, the result is reasonable. Due to the low amplitude of the grid-like features and the large dynamic range of coronal features, the visual effect of the flat field processing is not significant as that for the spot-like features. However, the images show that the radial structures in the corona, such as coronal plumes and coronal streamers, are clearer after the flat field correction. Therefore, we believe that the flat field processing greatly improves the quality of the EUV images for the coronal radiation observation of X-EUVI. 

\begin{figure}
\includegraphics[bb=30 0 502 560,width=1\textwidth,clip]{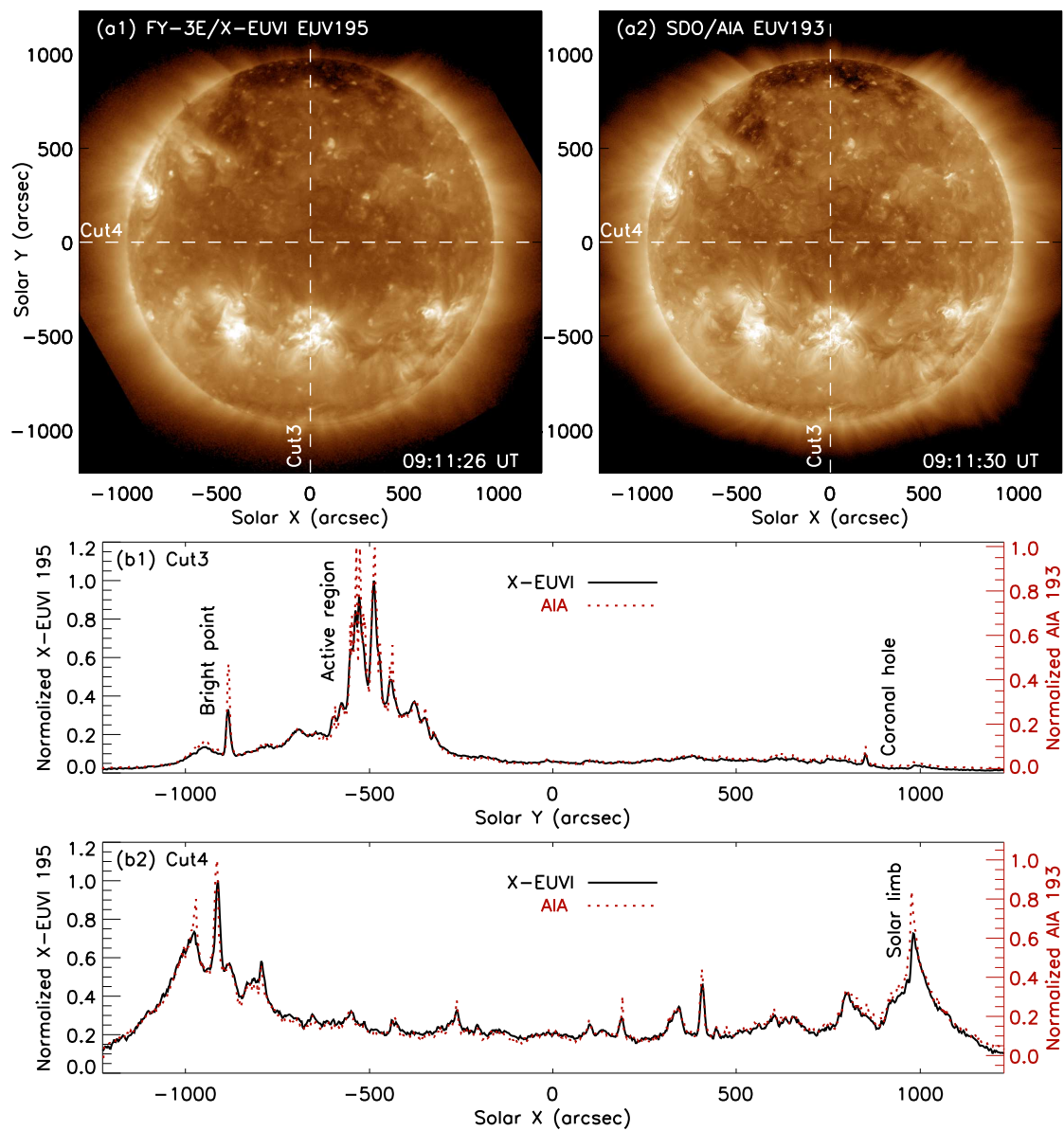}
\caption{Comparison of X-EUVI 195 {\AA} and SDO/AIA 193 {\AA} images on October 28, 2021. Cut3 is parallel to solar Y-axis, and Cut4 is parallel to solar X-axis. In panels (b1) and (b2), the black line represents X-EUVI 195 {\AA} and the red dotted line represents AIA 193 {\AA} light curves with normalized brightness,  respectively.}
\label{fig:fig6}
\end{figure}

After dark and flat field corrections with our method, the comparative analysis results show that the X-EUVI 195 {\AA} coronal radiation is consistent with that from the SDO/AIA 193 {\AA} channel. Figure \ref{fig:fig6}(a1) is an X-EUVI 195 {\AA} image at 09:11 UT on October 28, 2021, and Figure \ref{fig:fig6}(a2) is an almost simultaneous AIA 193 {\AA} image for comparison. The X-EUVI image has been aligned with the AIA image by the image registration process that includes image scale correction, shift and rotation corrections using the cross-correlation method manually.
It can be seen that the coronal features of the X-EUVI image on both solar disk and solar limb agree well with those of the AIA image. We take horizontal and vertical slices on the images, where Cut3 is parallel to solar Y-axis, and Cut4 is parallel to the solar X-axis. Figure \ref{fig:fig6}(b1) shows the brightness variation along Cut3, and features such as coronal bright points, active region loops, and a dark region corresponding to a coronal hole can be seen in this light curve. We normalized the brightness curve to facilitate comparative analysis, where the black line represents X-EUVI 195 {\AA} images and the red dotted line represents AIA 193 {\AA} data. The brightness variation along Cut4 shows limb brightening pattern in EUV wavelength (see Figure \ref{fig:fig6}(b2)). Although the curves of AIA 193 {\AA} channel are sharper at the coronal loops, the bright points, the solar limb, and other small-scale features because of its higher spatial resolution (0.{\arcsec}6 per pixel), the overall trends of FY-3E/X-EUVI 195 {\AA} and SDO/AIA 193 {\AA} light curves are consistent across the entire solar disk.

\section{Conclusion and Discussion}

The routine EUV 195 {\AA} imaging data from FY-3E/X-EUVI has imaging rotation with a speed of 3.5{\degr} per minute. If several offset solar images taken by changing the satellite’s pointing are used to derive the flat field, it will be very difficult because there are both rotation and shift during the image registration process. In the paper, we proposed a new post-hoc flat field measurement method taking advantage of the concentric rotation property in the routine data taken with the imaging stabilization system of the X-EUVI. An advantage of the post-hoc method is that it does not need to occupy the observation time for the calibration of flat field, which can facilitates convenient and continuous observations, and this is valuable for space weather operations.

The principle of our method is that the small-scale and time-varying component of the coronal radiation is approximately zero if we make a data cube from plenty of concentric rotation images and take the median value for each pixel along the time axis. The large-scale and invariable component can further be derived assuming that it is symmetrical under all rotations about the center of solar disk. Once the large-scale and invariable coronal radiation is removed, the remaining part is the flat field. For the X-EUVI's routine EUV 195 {\AA} imaging data, the flat field can be generated with cadences from hundreds minutes (one orbit) to several days. The accuracy of the flat field depends on how much coronal radiation is left with the limited data and the processing method. In the paper, the flat field is derived with the period of one orbit, one, two and three days. From the comparison, it is found that higher flat field accuracy can be achieved by employing more data. Obviously if more data (larger n) are used, it takes more time to generated a flat field. Once the instrumental flat field changes during this period, we merely obtained the averaged flat field. From the comparison of Figures \ref{fig:fig3}(a) and \ref{fig:fig3}(b), our method can monitor the changes in the flat field image due to the contamination on the CCD detector and other reasons, and thus it can support routine calibration. It provides a convenient flat field calibration option for observation instruments that are difficult to maintain manually, such as space-based instruments and automatic observatories located in remote areas. The signal to noise ratio from the original imaging data is not included in the derivation of flat field in section 2, which is discussed in \citet{Potts2008} and \citet{bai2018} and is the same for coronal observations.

The flat field from July 22, 2021 shows that the new method is able to measure the instrumental spot-like non-uniformity possibly from the contamination on the detector, which mostly disappear after the in-orbit self-cleaning process seen from that taken on October 29 (see Figures\ref{fig:fig3}(a) and \ref{fig:fig3}(b)). The dramatic change in the flat field images before and after the self-cleaning process also demonstrates the effectiveness of the self-cleaning mode of the X-EUVI. Furthermore, the quasi-periodic grid-like non-uniformity possibly from the obscuration of the support mesh on the rear filter can also be obtained from the flat field, which have fluctuation with a small amplitude and a period of about 30 pixels, and it is roughly in line with  the parameters of the support mesh. After the flat field correction, the instrumental non-uniformity from the original data is effectively removed. The narrow and bright ring at solar limb in the flat field image (see the blue arrow in Figure \ref{fig:fig3}(d)) possibly caused by limited instrument pointing accuracy and stability, which can lead to sub-pixel drift in the position of the center of solar disk, and sometimes resulting in slight asymmetry in the derived median image with Eq. \ref{eq5}. It may also be caused by the limited accuracy of the derived solar disk center. In future works, we will try to remove the influence of the ring and study the changes of the flat field on different time-scales. To better illustrate the availability of our flat field method, we compare the X-EUVI 195 {\AA} data after dark and flat fieldcorrections with the 193 {\AA}  imaging data from SDO/AIA. We found that they are consistent with each other except that AIA has higher spatial resolution. Overall, we provide a candidate method for the flat field measurement of current and future solar EUV telescopes.

\begin{acknowledgements}
We sincerely thank the in-orbit test and science teams of the National Satellite Meteorological Center for the FY-3E/X-EUVI data. The SDO data have been used by courtesy of AIA science teams. This work is supported by National Natural Science Foundation of China (41774195), Ten-thousand Talents Program of Jing-Song Wang, and the Specialized Research Fund for State Key Laboratories. X.Y.B. was supported by the Strategic Priority Research Program of the Chinese Academy of Sciences, Grant No. XDA 15018400. Z.Y.H. was supported by the China Postdoctoral Science Foundation (2021M700246).
\end{acknowledgements}

\bibliographystyle{raa}
\bibliography{ms2022-0218}
\label{lastpage}

\end{document}